\documentstyle[aps,prl,epsfig,12pt]{revtex}
\title{Constraints on the Existence of Strange Quark Stars} 
\author{Shmuel Balberg}
\address{Racah Institute of Physics, The Hebrew University, 
Jerusalem 91904, Israel} 
\date{\today}
\begin{document}
\setcounter{page}{1}
\maketitle
\vspace{0.25in}

%=================================================================
%    ABSTRACT
%=================================================================
\begin{abstract}
Creation of strange quark stars through strong interaction deconfinement 
is studied based on modern estimates of hyperon formation in neutron stars. 
The hyperon abundance is shown to be large enough so that if strange 
quark matter (SQM) is the true ground state of matter, 
the deconfinement density should be at most $2.5\!-\!3$ times the 
nuclear saturation density. If so, deconfinement occurs in neutron 
stars at birth, and {\it all} neutron stars must be strange quark 
stars. Alternatively, sould observation indicate that some neutron stars 
have a baryonic interior, SQM is unlikely to be absolutely stable.

\end{abstract}

\begin{center}
------------------------------
\end{center}

\vspace{0.2in}
PACS numbers: 97.60.Jd, 21.65.+f, 12.38.Mh
%Submitted to The Physical Review Letters

\newpage

One of the implications of the hypothesis that strange quark matter 
(SQM) is the true ground state of matter [1,2]
is that some or all of neutron stars are actually strange quark stars.
The properties of strange quark stars have been studied in many works [3-6],
and were found to be mostly similar to those of ``conventional'' neutron 
stars, where matter is in a baryonic phase. Recent studies of strange quark 
stars include unique cooling scenarios \cite{Schaabetal} and formation 
of strange dwarfs \cite{GlenSD}. 

A key issue regarding strange star formation is the conversion 
(deconfinement) process of baryonic matter to SQM. A direct consequence of 
the SQM hypothesis \cite{Witten} is that if a seed of SQM is
created in the interior of a neutron star, it will initiate a burning 
process that converts the entire star within a time scale of minutes 
[9-11]. It has been recently proposed that the emitted energy in such a 
conversion, typically estimated at several times $10^{52}$ ergs, could be a 
source of $\gamma$-ray bursts of cosmological origin \cite{ChengDai}. 

Several mechanisms for creating SQM seeds in neutron stars have been 
suggested \cite{Alcock86,Olin91}. These can be either internal, 
when the SQM seed is created 
in the star through hadron deconfinement, or external, when a SQM ``nugget'' 
which has been created elsewhere is absorbed in the star (or earlier in its 
progenitor). The abundance of free SQM nuggets in the universe 
(for example, from debri of strange stars which have coalesced with a binary 
counterpart) could be large \cite{FriedCall}, but is diffuclt to evaluate and 
initially depends on creation of strange quark stars through internal 
mechanisms. On the other hand, availability of 
internal mechanisms depends only on the structure of the neutron 
star. Furthermore, assuming the SQM hypothesis is correct, then if some 
internal mechanism of SQM seed formation is available in neutron stars 
at birth, all existing neutron stars must actually be strange 
quark stars. Alternatively, neutron stars can serve as a test of the SQM 
hypothesis: if a corrseponding internal mechanism is indeed available, but 
some of the observed neutron stars can be ruled out as being strange quark 
stars, the likely conclusion is that SQM is not the ground state of matter. 

The purpose of this Letter is to illuminate the role of hyperon
formation as an available mechanism for creating SQM in all neutron stars
{\it at birth}. A SQM seed can form within the neutron star if 
deconfinement of the hadronic matter is energetically favorable. 
Such deconfinement must proceed through the strong interaction, since multiple 
creation of strange quarks through the weak intearaction is suppressed. 
Hyperons, along perhaps with K$^-$ meson condensation, can initiate an 
internal mecahnism to create the SQM seed by providing the necassary 
strange quark content so that deconfinement proceeds through the strong 
interaction. While K$^-$ condensation is typically found to occur only at 
considerably higher densities of baryonic matter in equilibrium 
\cite{StonyK,SchafMish}, the consensus amoung recent works is that hyperon 
formation in neutron stars should begin at rather low densities. 
In fact, it is widely accepted [15-18] that hyperons begin to accumulate at 
about twice the nuclear saturation density, $\rho_0$ 
($\rho_0\approx0.16$ fm$^{-3}$). An example of the composition of neutron star 
matter in equilibrium, i.e., the fraction of each species as a function of 
the total baryon density, is given in Figure 1 (based on pevious work 
\cite{BalGal97}).

At low densities ($\rho_B\!\leq\!3\rho_0$) the relevant hyperons are the 
$\Sigma^-$ and $\Lambda$. Their combined abundance builds up a strangeness 
per baryon fraction, $|S|/A$, that exceeds $0.1$ at 
$\rho_B\!\approx\!2.5\rho_0$, and at $\rho_B\!\approx\!3\rho_0$ is already 
$\sim0.2$. These trends result from employing realistic 
hyperon$-$nuclear-matter interactions based on hypernucler data, and are 
weakly dependent on the model and the corresponding estimated equation 
of state. 
    
Contrary to claims in some previous works \cite{Olin91}, hyperon induced 
deconfinement does not require the baryon strangeness fraction to be equal to 
that of ground state SQM, $|S|/A\approx0.7\!-\!0.8$ \cite{FarJaf}. 
Rather, the condition is only that the phase transition into SQM with a
composition identical to that of the baryonic matter, e.g., strong 
deconfinement, be energetically favorable; the SQM then reaches its ground 
state by series of weak decays. 

It is often assumed that the strong interaction deconfinement requires the 
entire bulk to deconfine spontaneously into quark matter of equal composition 
and baryon number density (a discussion of a transition through 
an intermediate mixed phase follows below). In this case, the bulk baryon 
density, $\rho_B$, at which the transition is expected satisifies 
the condition 
\begin{equation} \label{eq:eqE/A}
  \frac{\varepsilon}{\rho_B}\left(\{b_i\},\rho_B\right)=
  \frac{\varepsilon}{\rho_B}\left(\{q_i(\{b_i\})\},\rho_B \right) \;\; ,
\end{equation} 

\setlength{\parindent}{0.0in}
where $\varepsilon$ is the energy density and $\varepsilon/\rho_B$ is the 
corresponding energy per baryon, $\{b_i\}$ denotes the baryon equilibrium 
composition and $q_i(\{b_i\})$ is the appropriate deconfined composition. If 
condition (\ref{eq:eqE/A}) is satisfied, a spontaneous strong interaction 
transition into SQM will occur.
Whether or not deconfinement will occur 
in a neutron star depends on the resulting density of deconfinement. If this 
density, which is dependent on the equations of state of the two phases, is 
reached in the star's interior, deconfinement is expected. 

\setlength{\parindent}{0.25in}
The quark matter equation of state may be evalulated within the MIT bag model 
detailed in \cite{FarJaf}. 
The specifics of the bag model are determined by the combinations of 
values for the bag constant, $B$, and the quark 
interaction coupling constant, $\alpha_c$. For each value of $\alpha_c$, 
$B$ is limited from below so that two-flavor quark matter 
is less bound than symmetric nuclear matter, since ordinary nuclei
do not deconfine strongly. The SQM hypothesis places an upper limit on $B$, 
in order for the ground state composition of SQM to be 
more bound than symmetric nuclear matter. 
Combinig these two conditions constrains the allowed range of values for the 
bag constant for any given value of  $\alpha_c$ which is consistent with the 
SQM hypothesis \cite{FarJaf}. For example, for $\alpha_c=0$, $B$ is 
limited to the range $56$ MeV fm$^{-3}\!\leq\!B\leq\!82$ MeV fm$^{-3}$. 

In order to estimate the density of deconfinement, energies per baryon were 
calculated for the baryonic and quark phases using various models of the 
baryonic equation of state from \cite{BalGal97}, and different bag 
models for the quark equation of state. As an example, Figure 2 compares 
the energy per baryon of baryonic matter with hyperons calculated with a 
model similar to $\delta\!=\!\gamma\!=\!\frac{5}{3}$ 
of \cite{BalGal97}, to the energy per baryon of quark matter with 
identical quark composition, using different combinations of $B$ and 
$\alpha_c$. The u and d quarks were assumed to be 
massless, and the mass of the s quark is set to 150 MeV. The transition 
density for each combination corresponds to the point where the energy per 
baryon in the baryonic phase crosses that of the quark phase.

The nonzero fraction of strange quarks lowers the energy per 
baryon in the deconfined phase, and it is found in this work 
that {\it all} models which predict that SQM is the true ground 
state also predict that spontaneous deconfinement should occur at 
densities lower than $3\rho_0$. It can be seen that a deconfinement 
density of $\rho_B\!\leq\!2.5\rho_0$ is found even for combinations 
of $B$ and $\alpha_c$ which are borderline for 
making SQM more bound than ordinary nuclear matter, i.e. 
($B\!=\!82$, $\alpha_c\!=\!0$), or ($B\!=\!63$, $\alpha_c\!=\!0.3$) 
\cite{FarJaf}. Even in the combination $(B\!=\!100,\alpha_c\!=\!0)$, 
for which the SQM hypothesis is no longer correct, the deconfinement 
density is still found to be lower than $3\rho_0$.

Quark matter models which do not allow spontaneous deconfinement are still 
possible, of course, like ($B\!=\!125$, $\alpha_c\!=\!0$) shown in Figure 2. 
In such models the corresponding binding energy of SQM is significantly larger 
than ordinary nuclear matter. In this case, some deconfinement might occur, 
and the resulting state could be coexisiting baryon and quark phases 
\cite{Glen2ph,Stony2ph}. 

The low values for the deconfinement density are mostly due to the 
presence of strange quarks. If hyperon formation is ignored, the baryonic 
matter includes only two flavors of quarks. It is found that nuclear matter 
in beta equilibrium may deconfine into two-flavor quark matter 
if quark matter is very bound ($B$ must be close to its lower limit for a 
given value of $\alpha_c$). For extreme quark matter models 
(such as $B\!=\!56$ MeV fm$^{-3}$, $\alpha_c\!=\!0$), deconfinement may 
occur even at densities of $\rho_B\!\approx\!\rho_0$, since nuclear matter at 
beta equilibrium has a higher energy per baryon than symmetric matter. 
However, other combinations of $B$ and $\alpha_c$ delay the deconfinement of 
nuclear matter to higher densities, and in some cases, quark matter does not 
form at any density.

As is expected, Similar results are found for other models of the baryon 
equation of state, since the equation of state of matter with hyperons 
is limited to a rather narrow range of values \cite{BalGal97}. In any case 
a valid equation must be stiff enough to support a maximum mass of at least 
1.4 $M_\odot$ (the determined mass of the 1913+16 pulsar).
%different quark matter models will also yield low density deconfinement 
While the quark bag model is clearly a simplified description of quark matter 
physics, it seems that the margin it allows for low density deconfinement into 
SQM is large enough, so that the qualitative results are unlikely to be 
dependent on the quark matter model as well.

The baryonic density at which the bulk deconfines may actually be regarded 
as an upper limit for the creation of quark matter. 
This may be demonstrated by considering an alternative scenario, where quark 
matter drops form in the matter through quasi-equilibrium combinations of 
coexisting baryonic and quark phases. 
Once finite size quark phase bubbles appear they act as a seed of SQM, 
which burns into its equilibrium composition and consumes the surrounding 
baryons through further weak interactions. 
 
Equilibrium between baryonic and quark phases 
invovles conservation of two charges (baryon number and 
electric charge), and so the phases need not have equal compositions nor 
equal densities. This has been pointed out by Glendenning 
\cite{Glen2ph} with respect to two phases in full equilibrium, and is also 
valid for quasi-equilbrium. The quasi-equilibrium conditions differ 
from those for two phases in full equilibrium \cite{Glen2ph,Stony2ph} 
since the initial deconfinement is assumed to be determined by the strong 
interaction alone (again, multiple creation of strange quarks is forbidden).
For matter with a given composition and total baryon number density 
the conditions for two-phase equilibrium are the 
appropriate Gibbs conditions for chemical and pressure equilibrium:
\begin{eqnarray} \label{eq:eqmu}
   \mu_n=2\mu_d+\mu_u\;\;\;,\;\;\;\mu_p=2\mu_u+\mu_d  \\
   \mu_\Lambda=\mu_d+\mu_u+\mu_s\;\;\;,\;\;\; \mu_{\Sigma^-}=2\mu_d+\mu_s  
   \nonumber 
\end{eqnarray}
and
\begin{equation} \label{eq:eqP}
   P_B=P_Q \;\;\; ,
\end{equation}
\setlength{\parindent}{0.0in}
where $\mu_i$ refers to the chemical potential of species $i$, and $P_B$ and 
$P_Q$ are the pressure in the baryonic and quark phase, respectively. Since 
weak interactions are ignored, full equilibrium between the species is 
not enforced, and each baryon species equilibriates with the quark phase 
seperately. Furthermore, during deconfinemnt the total 
number of quarks of each of the three species must remain constant. 
Deconfinement proceeds once equilibrium allows the quark phase to occupy 
a nonzero fraction of the volume. 
 
\setlength{\parindent}{0.25in} 
In this work deconfinement through quasi-equilibrium was found to occur at 
lower baryon densities than bulk deconfinement, mainly because the quark 
phase can have a differnet density and composition than the baryonic phase. 
For combinations of $B$ and $\alpha_c$ which enable SQM to be absolutely 
stable, coexisting baryonic and quark phase are found to appear even at 
$\rho_B\!=\!\rho_0$. This is in agreement with the results of 
\cite{Glen2ph,Stony2ph} regarding full equilibrium. In fact 
relaxing the condition of beta equilibrium in the quark phase yields even 
slightly lower densities of deconfinement than for full equilibrium. 
Appropriately, the bag model constants which prevent low density deconfinement 
are even further away from the range where SQM is absolutely stable: 
for example, for $\alpha_c\!=\!0$, only $B\!\approx\!200$ can delay 
deconfinement to $3\rho_0$. While quasi-equlibrium deconfinement might be 
suppressed (for example, due to finite-size effects), these results offer 
supprot to the main conclusion of the analysis of bulk deconfinement: 
if the SQM hypothesis is correct, deconfinement should occur at densities 
below $2.5\!-\!3\rho_0$.

The immdediate consequence of the above discussion is that in neutron stars 
with a central density greater or equal to $2.5\!-\!3\rho_0$, the
baryons should deconfine into SQM, if such matter is indeed absolutely stable.
The SQM will then convert through the weak interaction to its equilibrium 
composition, and proceed to convert the entire star into a strange quark star.

Most equations of state for high-density matter require a central density 
of $\rho_c\!\geq\!3\rho_0$ to support a mass of $1.3\!-\!1.5$ $M_{\odot}$, 
which is the range of observed neutron star masses. This is true even for 
equations which disregard hyperon formation (erroneously, acording 
to the above remarks), and is pronounced when hyperons are taken into account, 
since the inclusion of more baryon species softens the equation of 
state \cite{BalGal97,newSB}, calling for an even larger value of 
$\rho_c$. Hence, the most likely conclusion is that if the SQM 
hypothesis is correct, then all neutron stars should be strange quark stars. 
The conversion into strange quark matter will occur 
immediately at birth of the neutron star, perhaps after the initial 
neutrino-diffusion time. In any case, a ``delayed'' burning of a neutron 
star into a strange quark star later in its evolution seems to be ruled out. 

It can be argued that the nuclear equation of state might be 
stiff enough so that a star of $1.4M_\odot$ will have a central density lower 
than $2.5\rho_0$. Such equations cannot be excluded (and are sometimes found 
in relativistic mean field models, due to the reduced values of effective 
masses in these models), although this is inconsistent with most published 
equations of state.
However, a very stiff nuclear equation of state yields high values of the 
energy per baryon, and is thus susceptible to 2-flavor deconfinement. 
Only a very limited range of quark matter models 
(typically with $\alpha_c\!\approx\!0$ and appropriate relatively high 
values of $B$) can keep the nuclear matter in stiff equations confined up to 
$2.5\rho_0$, while still predicting that SQM is absolutely stable.
Furthermore, a stiff nuclear equation of state is also unfavorable 
in view of the current theory of core-collapse (type II) supernovae. 
 
Could all neutron stars be strange quark stars? The observation of glitch 
behaviour in pulsars severely limits this possibility. Glitches are sudden 
jumps in the rotation frequency of a pulsar, with a spin down rate of 
$\Delta\dot{\Omega}/\dot{\Omega}\sim 10^{-3}-10^{-2}$. Current models of the 
glitch phenomena rely on the neutron superfluid vortex creep theory 
(see \cite{PAglitch} for a review), which requires that the effective moment 
of inertia of the inner crust of the star, $I_i$, fulfill the condition 
$I_i/I\!\approx\!\Delta\dot{\Omega}/\dot{\Omega}$, where $I$ is the total 
moment of inertia of the star. This condition is typical of any 
two-component model for pulsar glitches. Following this analysis, Alpar 
\cite{Alpar} pointed out that this entire class of models for glitches must be 
discarded for strange quark stars, which are expected to have very 
small crusts ($I_i/I\sim10^{-5}$) \cite{Alcock86}. Glendenning and Weber have 
argued \cite{GlenWeb92} that glitches could originate even in the 
very low mass crust of strange stars, but up to date no model for 
strange quark star glitch has been suggested. Hence, it currently 
seems reasonable to conclude that at least glitching neutron stars are not 
strange quark stars. This argument has been used with regard to the 
possibility that the flux of strange quark nuggets in the universe is large 
enough to have converted all neutron stars to strange stars \cite{FriedCall}. 
In contrast, the present Letter points to the formation of 
hyperons as a more robust mechanism that could convert all neutron stars to 
strange quark stars, and is independent of the uncertainties in 
estimating the rate of SQM nugget production in binary coalescence.

In conclusion, it seems likely that all neutron stars should have central 
densities which allow for formation of a significant hyperon fraction. 
This result is basically model independent and suggests,
as demostrated above, a robust mechanism for the creation of strange quark 
matter in all neutron stars, if such matter is indeed the true ground state 
of baryonic matter. If this is the case, all neutron stars should convert
at birth to strange quark stars - a possibility which is difficult to combine 
with the lack of an explanation for the observed pulsar glitch phenomena. 
Hence, widely accepted evaluations of hyperon formation in neutron 
stars [15-18] serve as an indication that strange quark matter is not the 
true ground state of matter.

\vspace{0.16in}  
% Acknowledgements
The author is grateful to Avraham Gal for extensive guidance and for 
critically reviewing the manuscript. The author also wishes to thank Gideon 
Rakavi and Tsvi Piran for helpful discussion and comments. This research was 
partially supported by the U.S.-Israel 
Binational Science Foundation grant 94-68.

%========================================================================
\newpage
\begin{center}
{\large Figure Captions}
\end{center}
\setlength{\parskip}{0.1in}

Figure. 1. Equilibrium compositions for matter containing hyperons as 
well as nucleons and leptons, for a model similar to the model 
$\delta\!=\!\gamma\!=\!\frac{5}{3}$ described in \cite{BalGal97}. 

Figure. 2. Energy per baryon of baryonic matter in equilibrium, and the 
energy per baryon for quark matter of identical composition and density, as 
a function of the baryon number density. The energy per baryon in the 
baryonic phase (solid line) is calculated with the same model as in Figure 1, 
and the various curves of quark matter (dashed lines) correspond to 
bag models with different values of the bag constant, $B$, (in MeV fm$^{-3}$) 
and the coupling constant, $\alpha_c$, given in brackets as $(B,\alpha_c)$. 
The arrow marks the density where the first hyperons appear 
($\sim\!0.3$ fm$^{-3}$).

\newpage
\epsfig{file=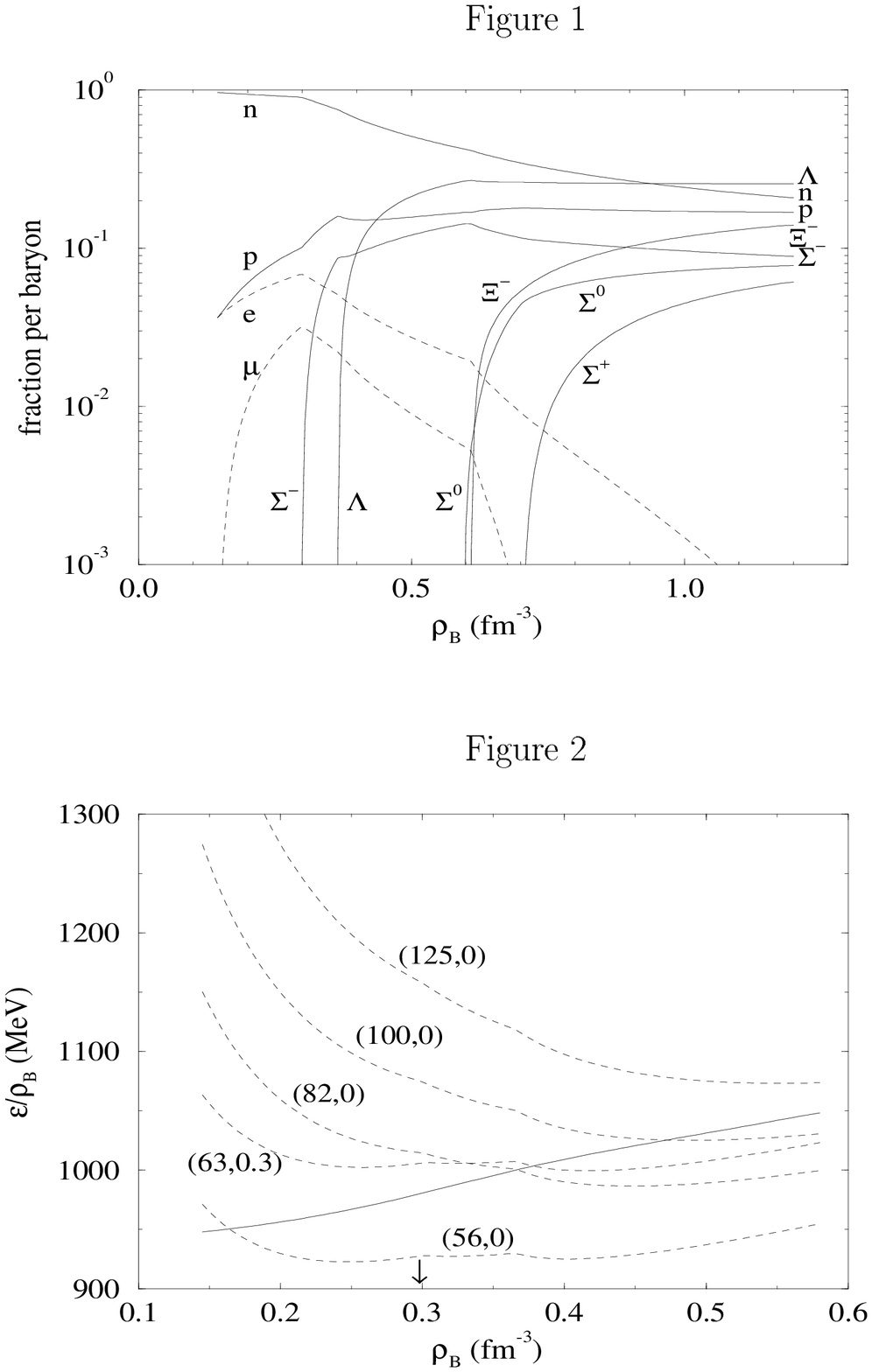,height=23.0cm,width=15.5cm}

\newpage
\begin{center}
{\large References}
\end{center}

\begin{thebibliography}{99}

\setlength{\parskip}{0.0in}

\bibitem{Witten} E. Witten,  
           Phys. Rev. D {\bf 30}, 272 (1984).

\bibitem{FarJaf} E. Farhi and R. L. Jaffe,
           Phys. Rev. D {\bf 30}, 2379 (1984).

\bibitem{Alcock86} C. Alcock, E. Farhi and A. Olinto,
           Astrophys. J. {\bf 310}, 261 (1986).

\bibitem{others1} P. Haensel, J. L. Zdunik and R. Schaeffer,
           Astron. and Astrophys. {\bf 160}, 121 (1986).  

\bibitem{others2} G. Baym, R. Jaffe, E. W. Kolb, L. McLerran and T. P. Walker,
           Phys. Lett. B {\bf 160}, 181 (1985).

\bibitem{Glenbook} For a review see  N. K. Glendenning,
           {\it Compact Stars}, (Springer, New York, 1996).

\bibitem{Schaabetal} C. Schaab, B. Hermann, F. Weber, 
                     and M. K. Wiegel, 
           Astrophys. J. Lett. {\bf 480}, L111 (1997).

\bibitem{GlenSD} N. K. Glendenning, C. Kettner and F. Weber,
           Phys. Rev. Lett. {\bf 74}, 3519 (1995);
                 N. K. Glendenning, C. Kettner and F. Weber,
           Astrophys. J. {\bf 450}, 253 (1995).

\bibitem{Olin87} A. V. Olinto, 
           Phys. Lett. B {\bf 192}, 71 (1987).

\bibitem{Olin91} A. V. Olinto,
           Nucl. Phys. {\bf B24} (Proc. Suppl.), 103 (1991).

\bibitem{ByPe}   H. Heiselberg, G. Baym and C. J. Pethick,
           Nucl. Phys. {\bf B24} (Proc. Suppl.), 144 (1991).

\bibitem{ChengDai} K. S. Cheng and Z. G. Dai,
           Phys. Rev. Lett. {\bf 77}, 1210 (1996).
    
\bibitem{FriedCall} R. R. Caldwell and J. L. Friedman,
           Phys. Lett. B {\bf 264}, 143 (1991).

\bibitem{StonyK}   V. Thorsson, M. Prakash and J.M. Lattimer, 
          Nucl. Phys.  {\bf A572}, 693 (1994).

\bibitem{SchafMish} J. Schaffner and I.N. Mishustin, 
          Phys. Rev. C {\bf 53}, 1416 (1996).                 

\bibitem{BalGal97} S. Balberg and A. Gal,
          Nucl. Phys. A (in press, 1997), 
          LANL preprint archive index nucl-th/9704013.

\bibitem{Glenn85} N.K. Glendenning, 
          Astrophys. J. {\bf 293} (1985) 470.	

\bibitem{newSB} P.J. Ellis, R. Knorren and M. Prakash, 
          Phys. Lett. B {\bf 349}, 11 (1995);
                R. Knorren, M. Prakash and P.J. Ellis, 
          Phys. Rev. C {\bf 52}, 3470 (1995).

\bibitem{Glen2ph} N. K. Glendenning,
          Phys. Rev. D {\bf 46}, 1274 (1992).

\bibitem{Stony2ph}  M. Prakash, J.R. Cooke and J. M. Lattimer,
          Phys. Rev. D {\bf 52}, 661 (1995).

\bibitem{PAglitch} D. Pines and M. A. Alpar,
          Nature {\bf 316}, 27 (1985).

                   M. A. Alpar, H. F. Chau, K. S. Cheng and D. Pines,
          Astrophys. J. {\bf 409}, 345 (1993).

\bibitem{Alpar}  M. A. Alpar,
           Phys. Rev. Lett. {\bf 58}, 2152 (1987).

\bibitem{GlenWeb92} N. K. Glendenning and F. Weber,
          Astrophys. J. {\bf 400}, 647 (1992).
\end{thebibliography}
\end{document}